\documentclass[aps,prl,reprint,groupedaddress]{revtex4-1}

\usepackage{natbib}
\usepackage[percent]{overpic}
\usepackage{xcolor}
\usepackage{amsmath}
\usepackage{hyperref}
\newcommand\solphys{Sol.~Phys.}%
\newcommand\aap{Astron.~Astrophys.}%
\newcommand\mnras{MNRAS}%

\graphicspath{{./}{./figs/}}
\begin{document}

\title{Modeling Repeatedly Flaring $\delta$ Sunspots}

\author{Piyali Chatterjee$^{1,2}$}
\email[]{piyali.chatterjee@astro.uio.no}
\author{Viggo Hansteen$^{1}$}
\author{Mats Carlsson$^{1}$}
\affiliation{$^{1}$Institute of Theoretical Astrophysics, University of Oslo, P.O. Box 1029 Blindern, N-0315 Oslo, Norway}
\affiliation{$^{2}$Indian Institute of Astrophysics, II Block Koramangala, Bengaluru-560034, India}
\date{\today}

\begin{abstract}
Active regions (AR) appearing on the surface of the Sun are classified into $\alpha$, $\beta$, $\gamma$, 
and $\delta$ by the rules of the Mount Wilson Observatory, California on the basis of their topological complexity. 
Amongst these, the $\delta$-sunspots are known to be super-active and produce the
most X-ray flares. Here, we present results from a simulation of the Sun by mimicking the upper layers and the corona, but 
starting at a more primitive stage than any earlier treatment. We find that this initial state consisting of only a thin 
sub-photospheric magnetic sheet breaks into multiple flux-tubes which evolve into a colliding-merging system of spots of opposite 
polarity upon surface emergence, similar to those often seen on the Sun. The simulation goes on to produce many exotic $\delta$-sunspot 
associated phenomena: repeated flaring in the range of typical solar flare energy release and ejective helical flux 
ropes with embedded cool-dense plasma filaments resembling solar coronal mass ejections.
\end{abstract}

\pacs{{96.60.Hv}, {96.60.Iv}, {96.60.ph}, {96.60.qd}, {96.60.qe}, {96.60.qf}}

\maketitle

Delta-sunspots are formed when two sunspots of 
opposite polarity magnetic field appear very close to each other and reside in the same penumbra, the radial 
filamentary structure outside the umbral region of the strongest magnetic fields. Strong shear and horizontal magnetic fields 
often exist at the polarity-inversion line separating the two polarities \citep{Zirin1988}. 
The subsurface processes which form the $\delta$-sunspots are still debated. Early observational studies \citep{Tang1983, Zhongxian_Wang94} propose that $\delta$-sunspots form from collision-merging of topologically separate dipoles, while 
numerical simulations by 
\cite{Linton_etal1999, Takasao_etal2015} 
show that kink unstable magnetic flux-tube -- 
helical field lines winding around a central axis -- emerging from the subsurface can have a $\delta$-sunspot like structure. 
More recently, attempts to model the $\delta$-spot in the 
NOAA AR 11158 utilized a uniformly twisted sub-surface flux-tube initially buoyant in two adjacent 
regions along its length \citep{Fang_Fan2015, Toriumi_etal2014}. Also, \cite{Mitra_etal2014} found a magnetic flux concentration 
resembling a $\delta$ sunspot in their stratified helical dynamo simulation. {These studies did not report any flaring activity. 
On the other hand \cite{Archontis_Hood2008} initialized their simulation with two parallel flux-tubes each 
lying at a different depth from the surface and with a different value of the initial magnetic twist 
which later evolved into a $\delta$-sunspot like structure and powered multiple reconnection events.}
Delta-sunspots are highly flare-productive -- 95\% of the strongest (X-class) X-ray flares originate from these 
regions \citep{Zhongxian_Wang94}. 
Using a realistic numerical simulation \cite{Archontis_Hansteen2014} showed that interaction between 
adjacent expanding magnetic bipoles pressing against each other can lead to the formation of strong current 
layers in the atmosphere which in turn lead to repeated flaring in 
the region. Here, we report on a three-dimensional magneto-hydrodynamic (MHD) simulation of the 
formation of a $\delta$-sunpot 
like region as a result of the break-up of a cool magnetic layer embedded in the upper 
solar convection zone into several flux-tubes
due to the growth of three-dimensional unstable modes excited in the layer. This instability is well known in the
literature as the undular instability (UI) \citep{Acheson1979, Fan2001}. The magnetic layer initially inserted is 
thinner than the local pressure scale height -- a precondition for UI. A detailed study of UI was performed 
by \citep{Brandenburg_Schmitt1998, Chatterjee_etal2011} in a very similar setup but 
without convection. They also calculated the amount of magnetic twist that is generated 
by the UI inside a thin magnetic layer with zero initial twist in a stratified, and rotating plasma. 
The latter study also found that the tubes formed 
are more twisted with increasing rotation 
Also, the sign of the mean twist in the domain changes with the sense of the rotation vector. {As it is not yet possible to observationally discern the sub-surface structure of the sunspots, most simulations so far employ
only uniformly twisted kink-unstable cylindrical flux tubes as the initial condition. The
amount of twist applied is a free parameter and so are the segments where the tubes must be initially buoyant.
Our simple initial condition alleviates the need for 
such free parameters.}

We solve the equations of compressible 
magneto-hydrodynamics  
in a $36 $ Mm $\times$ $36 $ Mm $\times$ $25 $ Mm Cartesian box using the higher-order 
finite difference code, 
the Pencil Code\footnote{{https://github.com/pencil-code/}}. 
The box rotates with a solar-like angular velocity 
$\Omega = 2.59\times10^{-6}$ s$^{-1}$ 
making an angle of $30^o$ with the 
vertical $\textrm{z}$-direction.
The box is resolved using a uniformly spaced grid 
with $d\textrm{x} = d\textrm{y} = 96$ km and $d\textrm{z}=48$ km. 
The initial state is a convectively relaxed state
with the vertical profiles of density, $\rho$, and temperature, $T$, 
given by Figure~\ref{fig0} (a). 
The domain consists of a sub-photospheric 
super-adiabatic layer in the lower 8.5 Mm of the box. The layer above ($0 < \textrm{z} < 2$ Mm) is cooled by a 
radiative cooling term $\propto \rho^2\Lambda(T)$ in the entropy equation to drive the surface convection and mimic the 
photosphere and the chromosphere. The temperature in this layer connects via a transition region to an 
isothermal corona ($3.5 ~\textrm{Mm} < \textrm{z} < 16.5 ~\textrm{Mm}$) maintained at $8.0 \times 10^5$ K by a Newtonian 
{cooling} term.
Into this steady state atmosphere we introduce a horizontal magnetic sheet at $\textrm{z}_0=-7.75 ~\textrm{Mm}$ with the
magnetic field vector, $\bf{B}$, strongly oriented in the x-direction as shown in Figure~\ref{fig0} (b). 
The horizontal extent of the sheet is 
about $-3~\mathrm{Mm} < \mathrm{y} < 3~\mathrm{Mm}$ and the maximum half-width, 
$R$, is 0.3 Mm at $\textrm{y}=0$. The sheet is neutrally buoyant with respect to the surroundings to prevent
it from rising immediately. However, for maintaining magneto-static equilibrium we cool it 
by reducing the specific entropy inside the sheet. 
The initial plasma-$\beta$, which is the 
ratio of gas pressure to magnetic pressure is $\sim 0.6$.
Simulations of the generation of such a magnetic sheet due to 
dynamo action and its subsequent break-up into 
buoyant flux-tubes have been performed \cite{Guerrero_Kapyla2011}. 
{It is also likely that the apex of an $\Omega$-shaped
ribbon rising from deep inside the convection zone as in Figure~3 (b) of 
\cite{Nelson_etal2011} can be described in terms of such a sheet.}
\begin{figure*}
\label{fig0}
\begin{overpic}[width=0.43\textwidth]{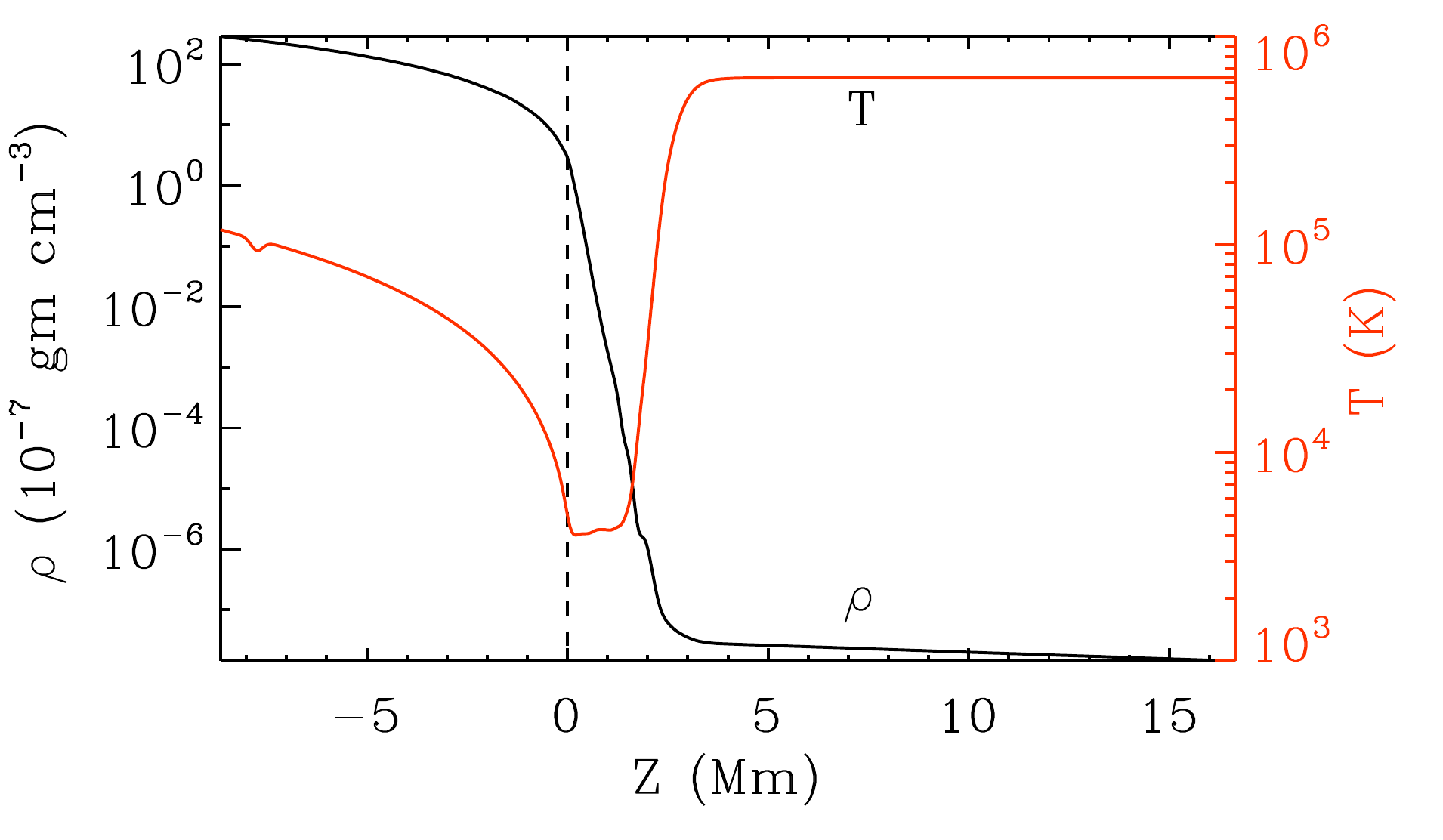}
\put(21,45) {\small (a)}
\put(16,33) {$\uparrow$}
\end{overpic}
\begin{overpic}[width=0.43\textwidth]{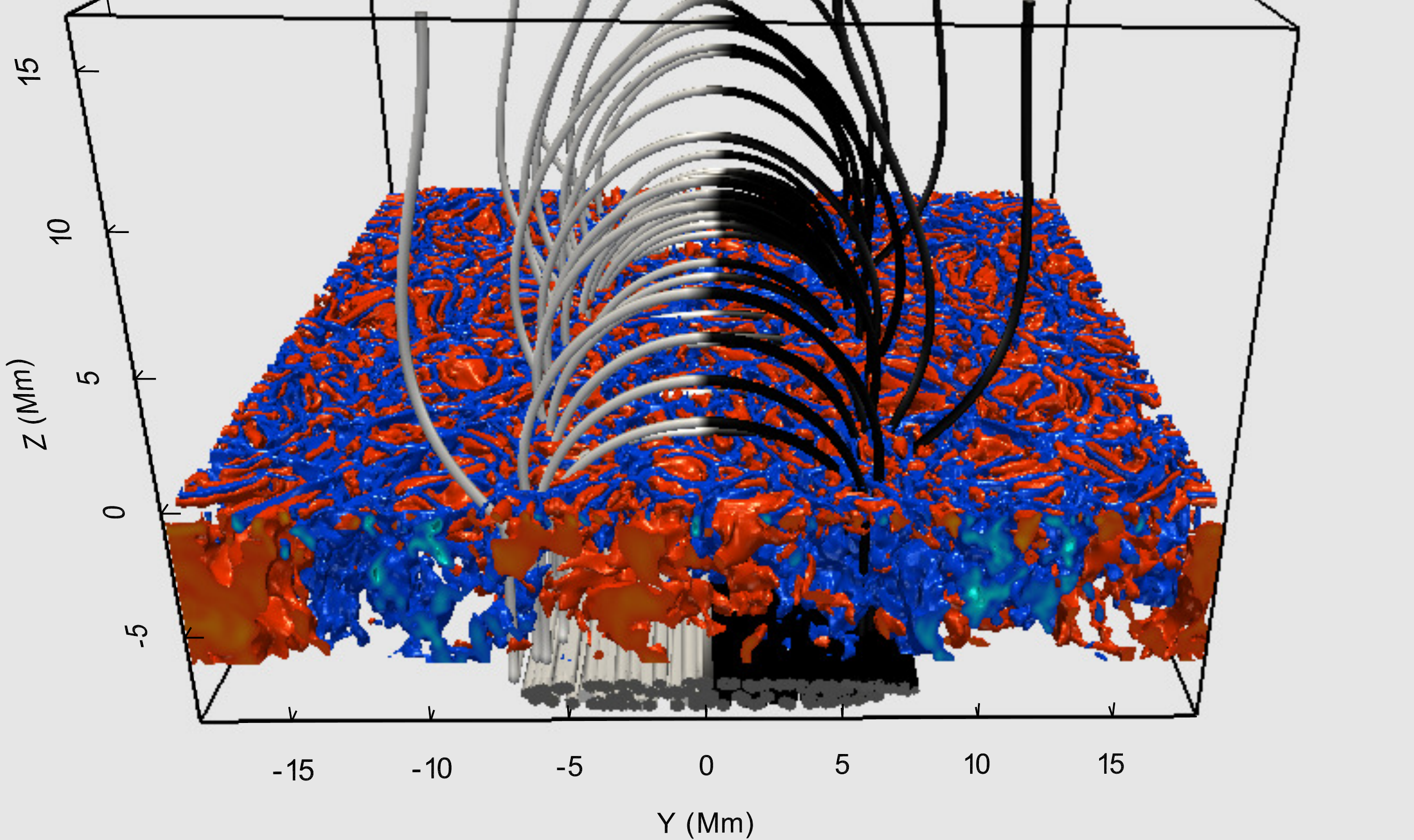}
\put(21,52) {\small (b)}
\end{overpic}
\caption{\label{fig0} (a) Vertical profiles of the horizontal averages of 
initial density and temperature. The arrow indicates the 
position of the cool magnetic sheet. (b) Initial magnetic field-lines shaded 
black (white) indicating positive (negative) $B_\textrm{z}$
The iso-surfaces of the upward (red) and the downward (blue)
velocity in the convective layer highlights the granulation pattern.}
\end{figure*}
Furthermore, we introduce an ambient magnetic field in the form of a 
potential field arcade at $\mathrm{z}> 0$ also shown in Figure~\ref{fig0}(b)
to facilitate the perturbations formed due to the
insertion of the strong magnetic sheet to 
travel along the field lines and out of the domain. 
Initially, the magnetic field at the photospheric foot points of the parallel arcade is 10 G. 
The lower boundary at $\textrm{z}=-8.5$ Mm is closed whereas the upper 
boundary only allows mass outflow with a vertical magnetic field condition. The x-boundaries are periodic, 
while the y-boundaries are perfectly conducting walls. 
Further details about the MHD equations, initial conditions, and the dissipation coefficients used here
are given in \footnote{See Supplemental Material at [URL] for model details}.
\begin{figure*}
\label{fig1}
\begin{overpic}[width=0.434\textwidth]{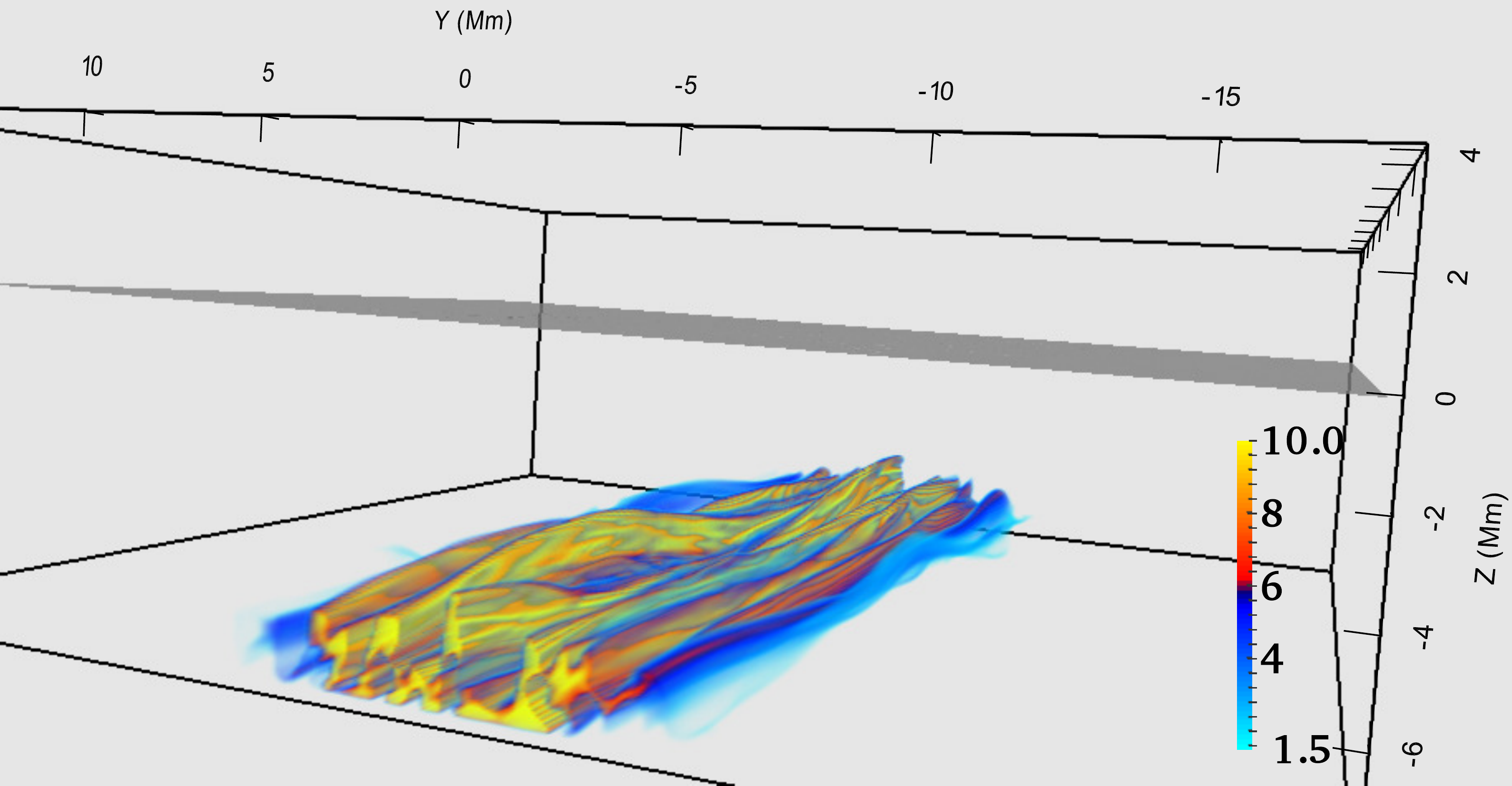}
\put(5,35) {\small (a) $t=33.0$ min}
\end{overpic}
\begin{overpic}[width=0.434\textwidth]{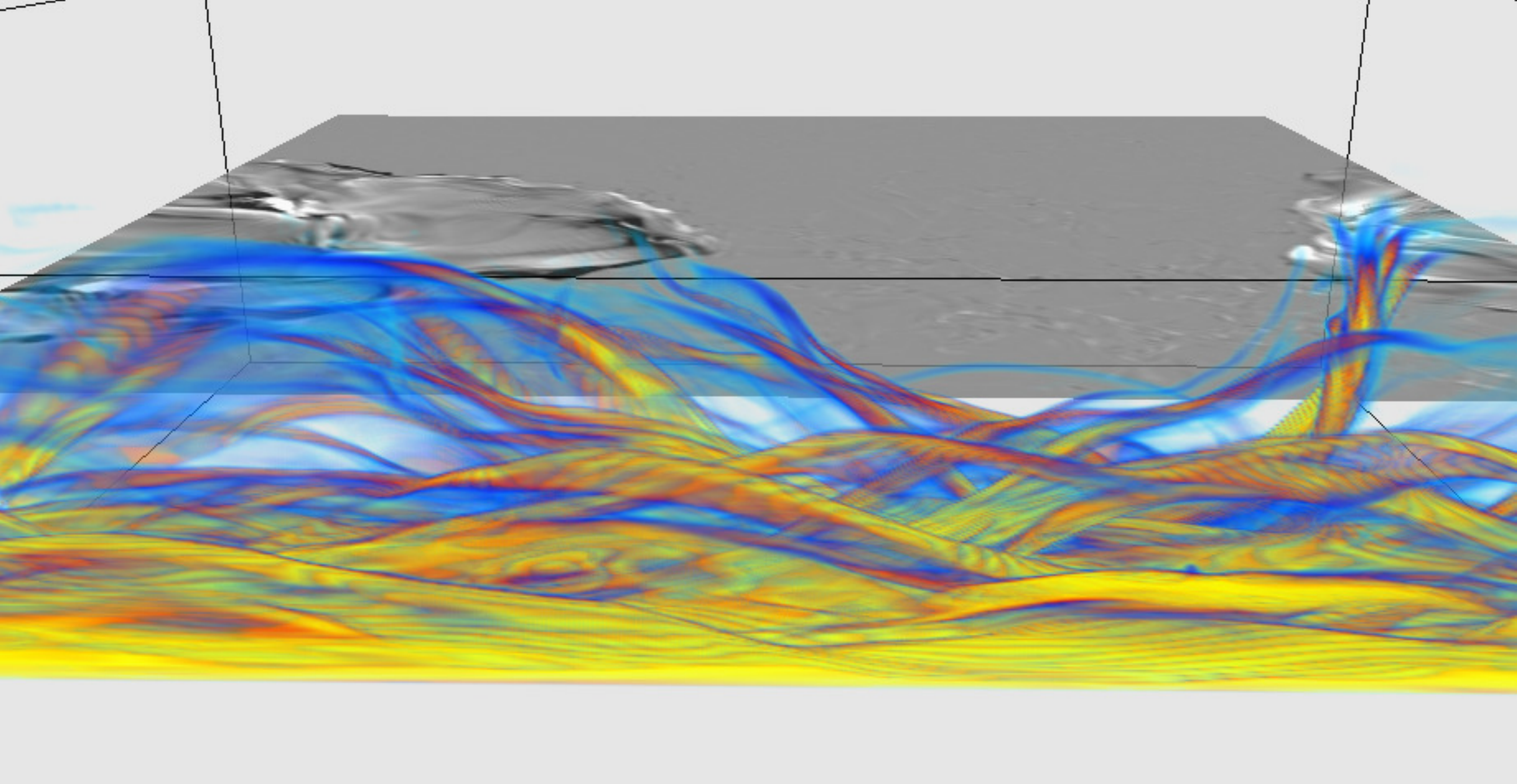}
\put(5,40) {(b) $t=112.2$ min}
\put(40,8) {$\xrightarrow[\small{\textrm{X}}]{}$}
\end{overpic}
\caption{\label{fig1} Volume rendering of the quantity $B\rho^{-1/4}$.
The gray shaded surface represents $B_\textrm{z}$ in the $\textrm{z}=0$ layer.} %
\end{figure*}

Our numerical simulation 
was run for 263 min of solar time.
We show the breaking up of the magnetic sheet into tubes 
and its subsequent evolution in Figure~\ref{fig1}a-b using volume rendering of a scalar quantity $B\rho^{-1/4}$, 
where $B$ is the magnetic field strength.  Here, the subsurface convection excites several spatial 
modes at the onset of the UI. As a result, the wave number along $\mathrm{x}$ is between one and two and the 
wave number in $\mathrm{y}$ is larger than twelve, which is the number of tubes counted from Panel (b). 
{The dominant mode excited will likely depend on the aspect ratio of the initial magnetic sheet 
as well as on the boundary conditions in the $\mathrm{x}$-direction.}
The magnetic field strength 
inside the newly formed flux tubes at $\mathrm{z}=-7.75$ Mm is about 25 kG. Figure~\ref{fig1} (b) depicts the progenitors of the $\delta$-sunspot as separate fronts of
positive and negative $B_{\mathrm{z}}$ emerging at $\mathrm{z}=0$ which successively move closer. 
The progenitors consist of different flux systems 
even though we traced some field lines directly connecting 
the spots below the photosphere. 
\begin{figure*}
\label{fig2}
\begin{overpic}[width=0.9\textwidth]{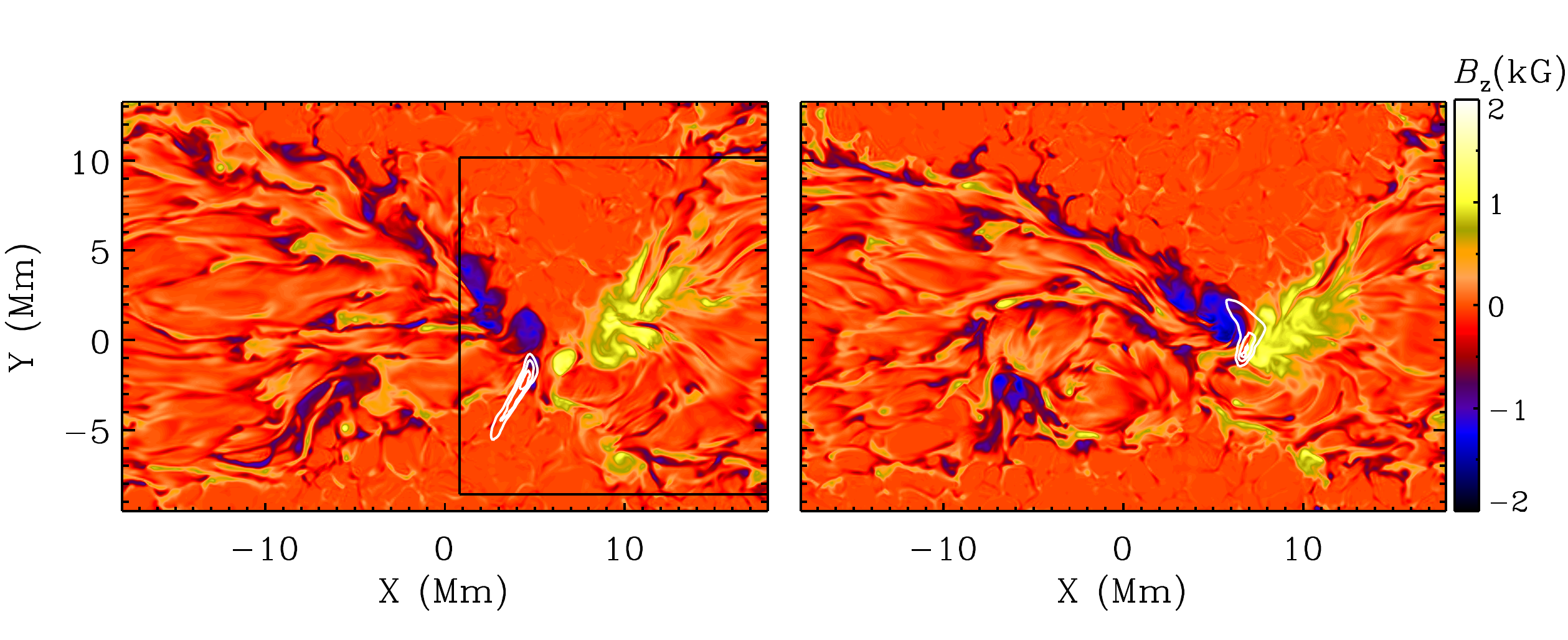}
\put(15,40) {\color{white}(a) $t=168.9$ min}
\put(55,40) {\color{white}(b) $t=198.9$ min}
\end{overpic}
\caption{\label{fig2} Vertical component of the magnetic field, 
$B_{\mathrm z}$, at $\mathrm{z}=0$ (shaded contours) at times indicated. 
Overlaid line contours (white) in (a) represent temperature contour 
levels at $\mathrm{z}=3.26$ Mm for values 90,000K, 1MK, 1.2MK whereas the 
respective temperature contour levels in (b) are at 90,000K, 1MK, 2.2MK. This figure is 
available as an animation at
\url{http://www.mn.uio.no/astro/english/people/aca/piyali/fluxemerge/fig3prl.mp4}.}
\end{figure*}
Figure~\ref{fig2} shows snapshots of the photospheric 
vertical magnetic field before and after the formation of the $\delta$-sunspot region upon collision of the 
opposite polarities of similar sizes and magnetic field strength. The unsigned magnetic flux emerging into the 
black square reaches a maximum of $7.25\times 10^{20}$ Mx at $t=200$ min and decreases slightly thereafter. 
The thread-like 
patterns of mixed polarities seen on both sides of the $\delta$-sunspot in the photospheric magnetogram 
indicate that there are two emerging and expanding bipolar regions side-by-side, the lateral extremes of 
which may be imagined to go beyond the periodic $\mathrm{x}$-boundaries.The collision is marked by several flares. 
A solar flare occurs when
magnetic energy is suddenly released in the form of heat, radiation, and energetic particle emission
inside very thin current sheets -- that are regions of very strong magnetic field gradients and thus sites 
where the magnetic field topology undergoes a major change. These flares thus reconnect field lines, 
ultimately leading to the components of the $\delta$-sunspot
becoming increasingly directly connected by field lines above the region. This prevents the component polarities 
of the $\delta$-sunspot from separating during the rest of the evolution.  Out of several flares we have been able to isolate only two strong ones, which
can be located in Figure~\ref{fig2} a,b by the white coloured contours of temperature at $\mathrm{z}=3.25$ Mm. 
The plasma is heated to a maximum temperature of $2.5$ MK at this height where the average temperature is 54,000 K. 
The onset of the two flares can also be identified as 
the locations of peaks (dashed lines) in the temporal evolution of the 
magnetic energy, $\mathcal{E}_B$ (dashed-dotted line in Figure~\ref{fig3}(a)), 
inside a sub volume of the domain with $0 < \mathrm{z} < 16.5$ Mm  and 
the horizontal extent demarcated by a black square in Figure~\ref{fig2}(a). 
The flares are powered by the magnetic energy
transported from the convection zone to the solar atmosphere.
\begin{figure}
\label{fig3}
\begin{overpic}[width=0.45\textwidth,clip=true]{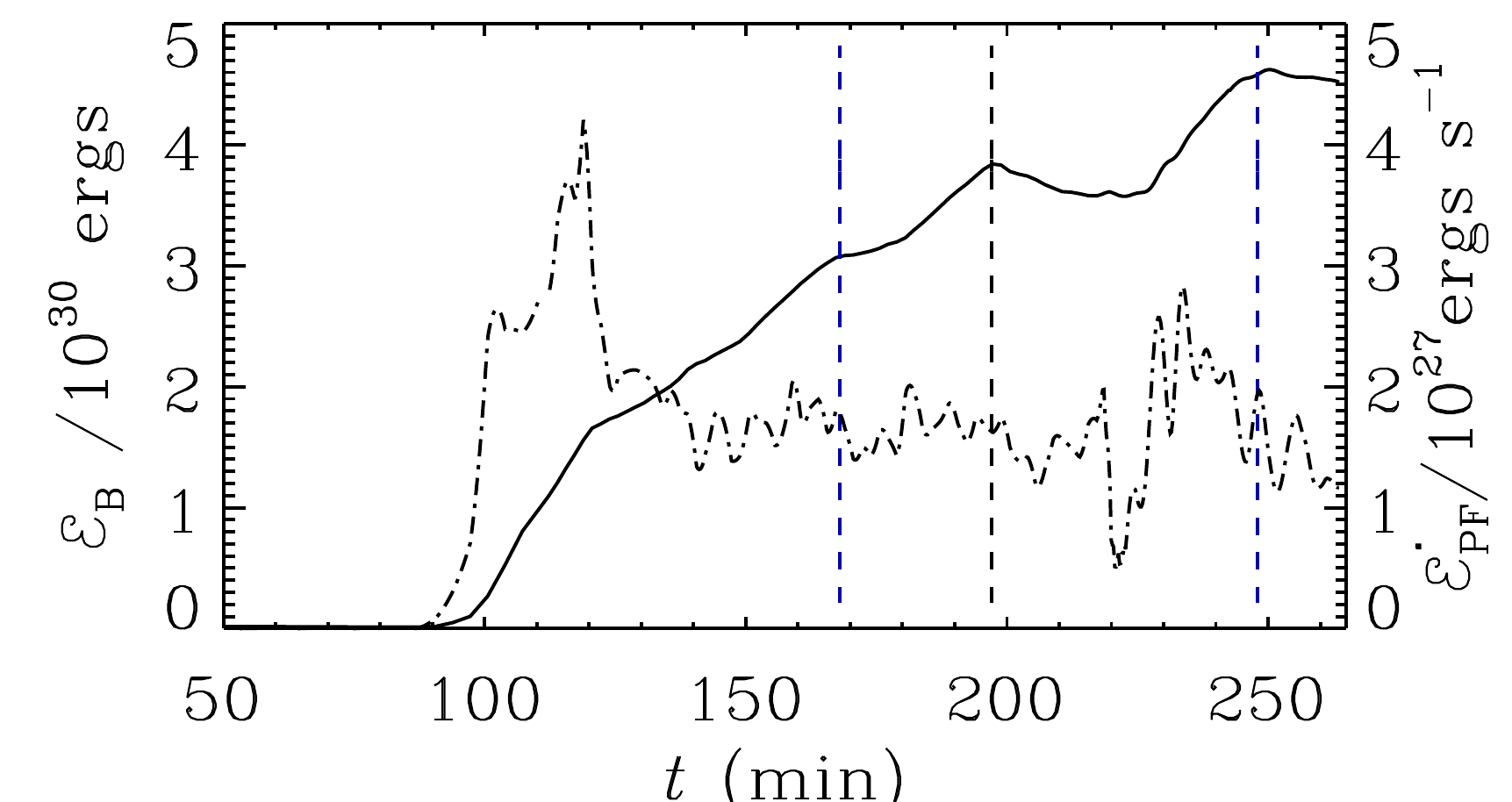}
\put(21,45) {(a)}
\end{overpic}
\begin{overpic}[width=0.43\textwidth]{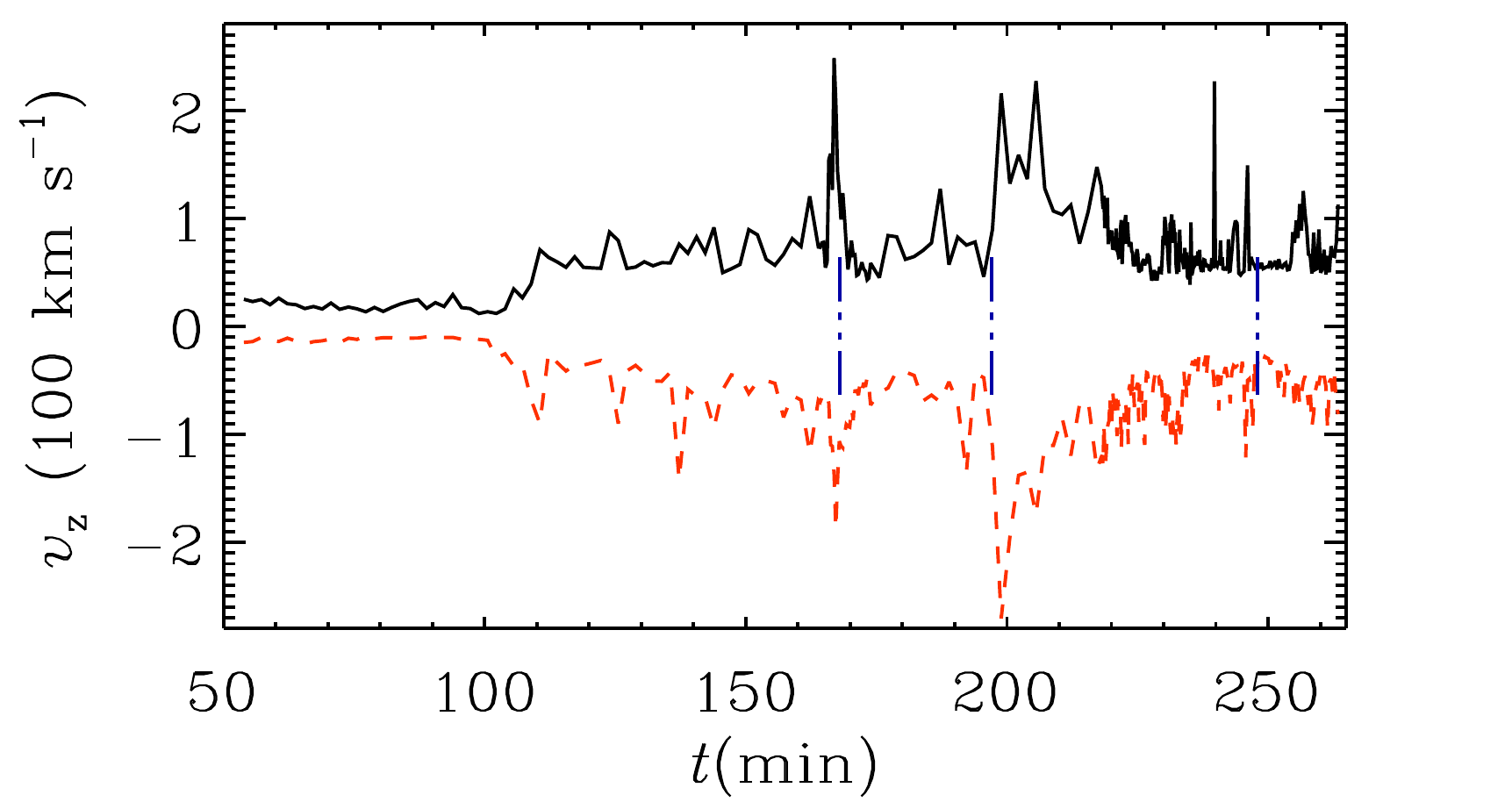}
\put(21,47) {(b)}
\end{overpic}
\caption{\label{fig3}Evolution of (a) magnetic energy, $\mathcal{E}_B$ (solid),  and $\dot{\mathcal{E}}_{PF}$ 
(dashed-dotted) and that of (b) maximum positive (solid) and negative values (dashed) of vertical velocity 
above the active region. The three vertical lines denote the times of the two flares and a flux rope eruption respectively.} 
\end{figure}
The rate of the magnetic energy input, or the 
Poynting flux, integrated over the faces of the same sub volume is,
$\dot{\mathcal{E}}_{PF} = \int c{\bf{E}}\times{\bf{B}}\cdot d{\bf{S}}/4\pi$ where, $c$ is the speed of light, 
$\bf{E}$ is the electric field vector, 
and $d\bf{S}$ is the area element. 
The time evolution of $\dot{\mathcal{E}}_{PF}$ is shown by the dashed-dotted line in Figure~\ref{fig3} (a).
The maximum possible flare energy, $E_{\textrm{\tiny flare}}^{\textrm{\tiny max}}$
is related to the difference in magnetic energy, $\Delta \mathcal{E}_B$, after and before the flare as,
\begin{displaymath}
\Delta \mathcal{E}_B=-E_{\textrm{\tiny flare}}^{\textrm{\tiny max}}+\int_{\Delta t_{\textrm{\tiny flare}}}\dot{\mathcal{E}}_{PF}dt,
\end{displaymath}
with $\Delta t_{\textrm{\tiny flare}}$ being the duration of the flare. From Figure~\ref{fig3} as well as 
the animation of Figure~\ref{fig2}, the flare occuring at $t=167.5$ min lasts for 5 min 
and the one at $t=197.2$ min lasts for 25 min. 
We estimate the magnetic 
energy release, $E_{\textrm{\tiny flare}}^{\textrm{\tiny max}}$ in the two cases to be $3.3\times10^{29}$ ergs 
and $1.7 \times 10^{30}$ ergs, respectively. The rate of energy release amounts to $1.1\times10^{27}$ ergs s$^{-1}$ 
in both cases, which agrees very well with the estimate made by \cite{Isobe_etal2005} for a C-class flare that occurred 
on November 16, 2000.
The magnetic energy dip at $t=240.2$ min is due to the 
eruption of a highly twisted flux rope releasing at least $2.3\times10^{30}$ ergs. We also observe 
good temporal correlation between the onset of energy release and bipolar reconnection jets appearing as pairs of 
maximum negative and positive values ($\sim \pm 270$ km s$^{-1}$) of vertical velocity (panel (b)).

\begin{figure}
\label{fig4}
\raggedright{\begin{overpic}[width=0.21\textwidth,clip=true]{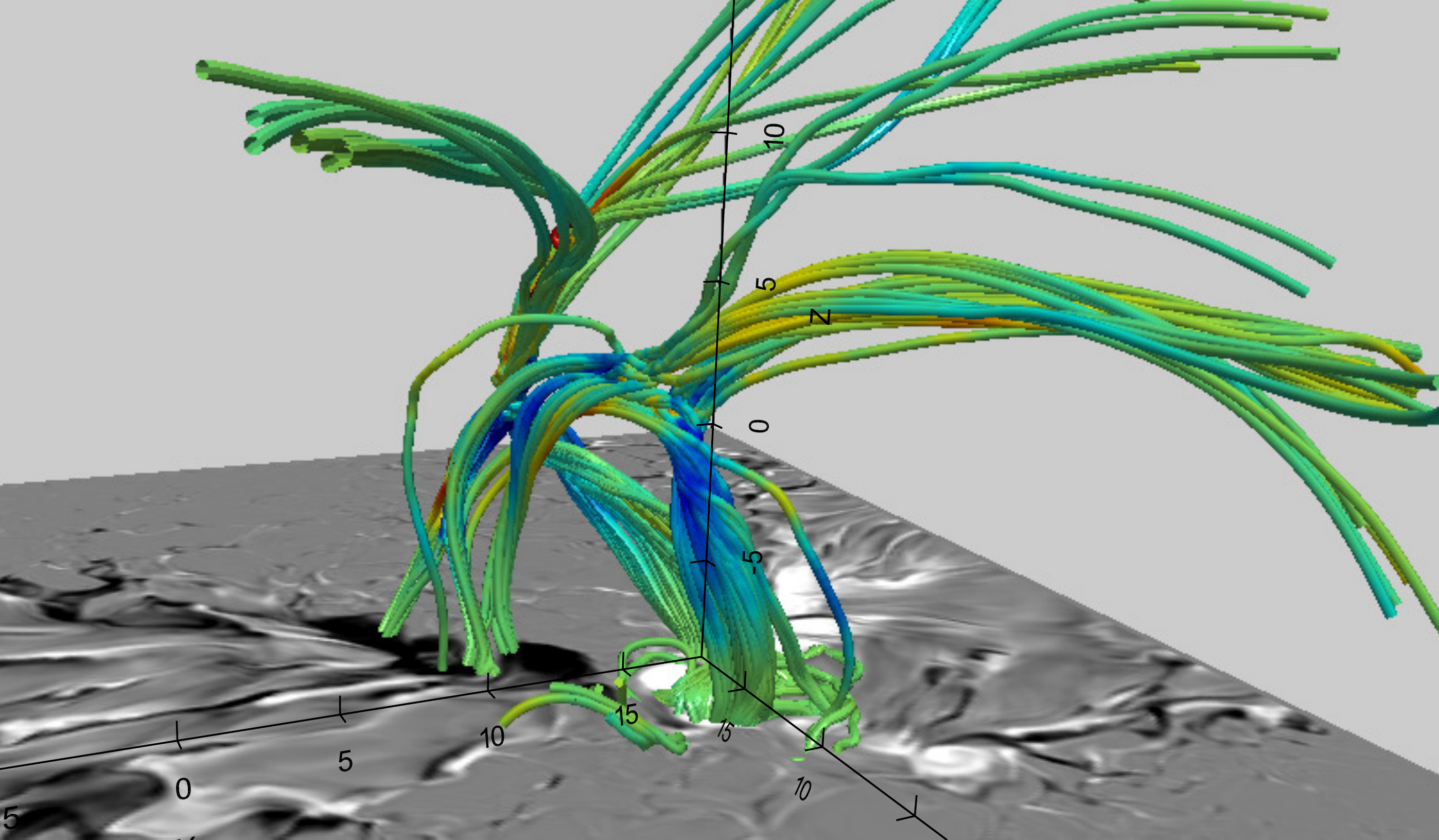}
\put(5,50) {(a) $t=168.8$ min}
\put(10,30){\small{FR-1}}
\put(100,-48){\includegraphics[width=0.245\textwidth,clip=true]{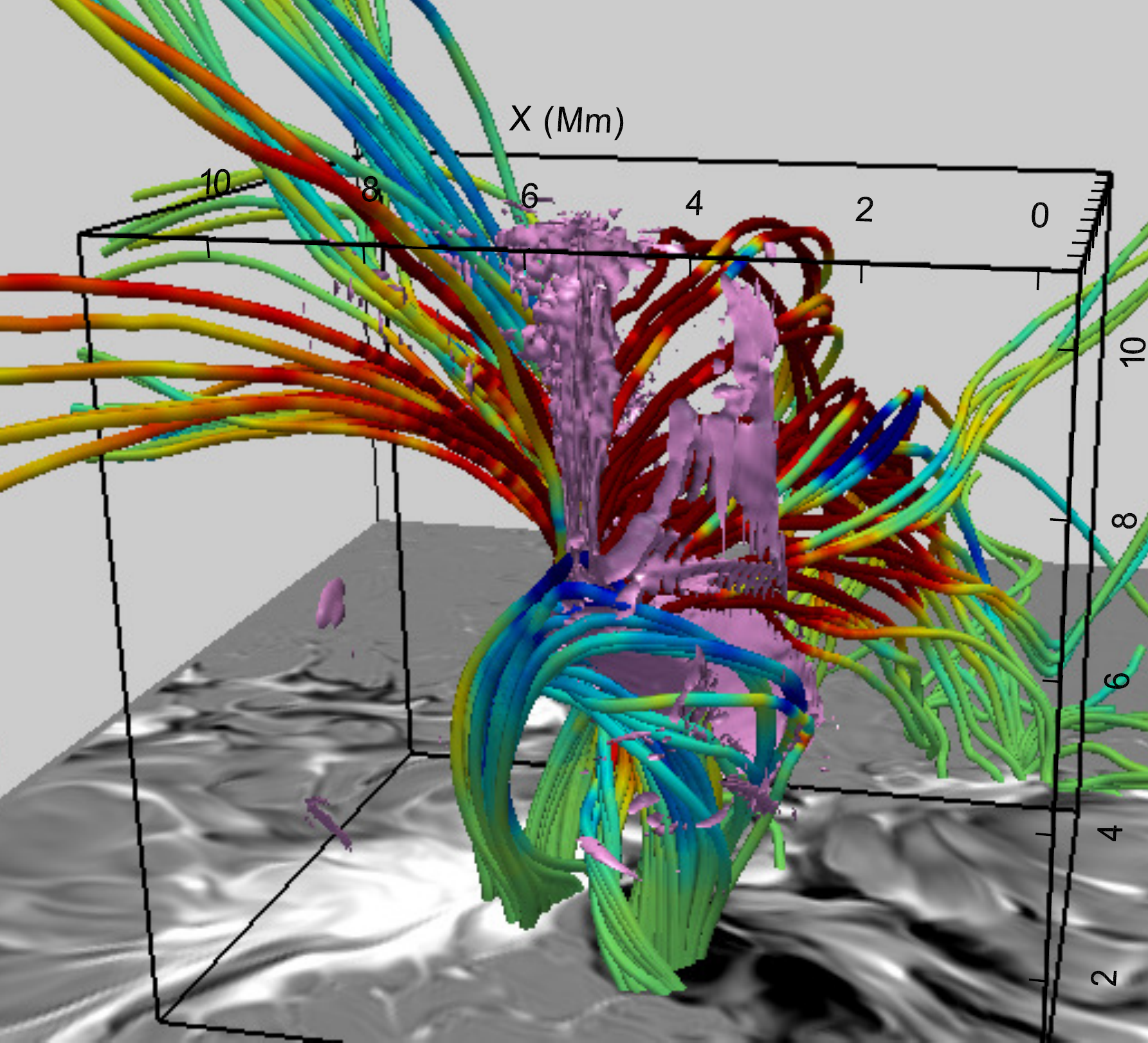}}
\put(150,50){(c) $t=198.9$ min}
\put(170,10){\small\color{white}{EFR-1}}
\put(125,-5){\small\color{white}{FR-2}}
\end{overpic}\\
\begin{overpic}[width=0.21\textwidth,clip=true]{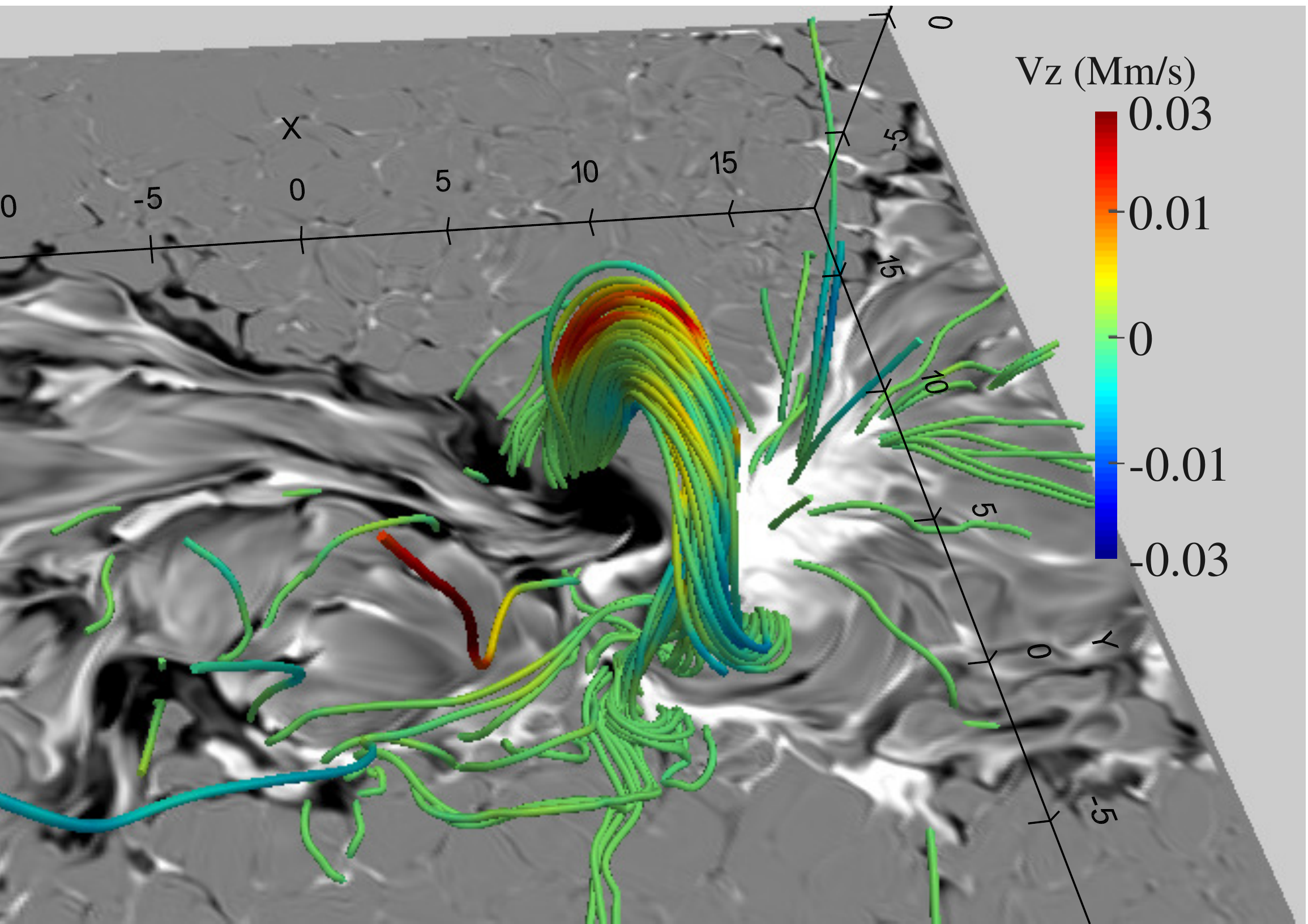}
\put(5,55) {\color{white}(b) $t=193.2$ min}
\put(60,40){\color{white}{\small{FR-1}}}
\end{overpic}}\\
{\begin{overpic}[width=0.21\textwidth,clip=true]{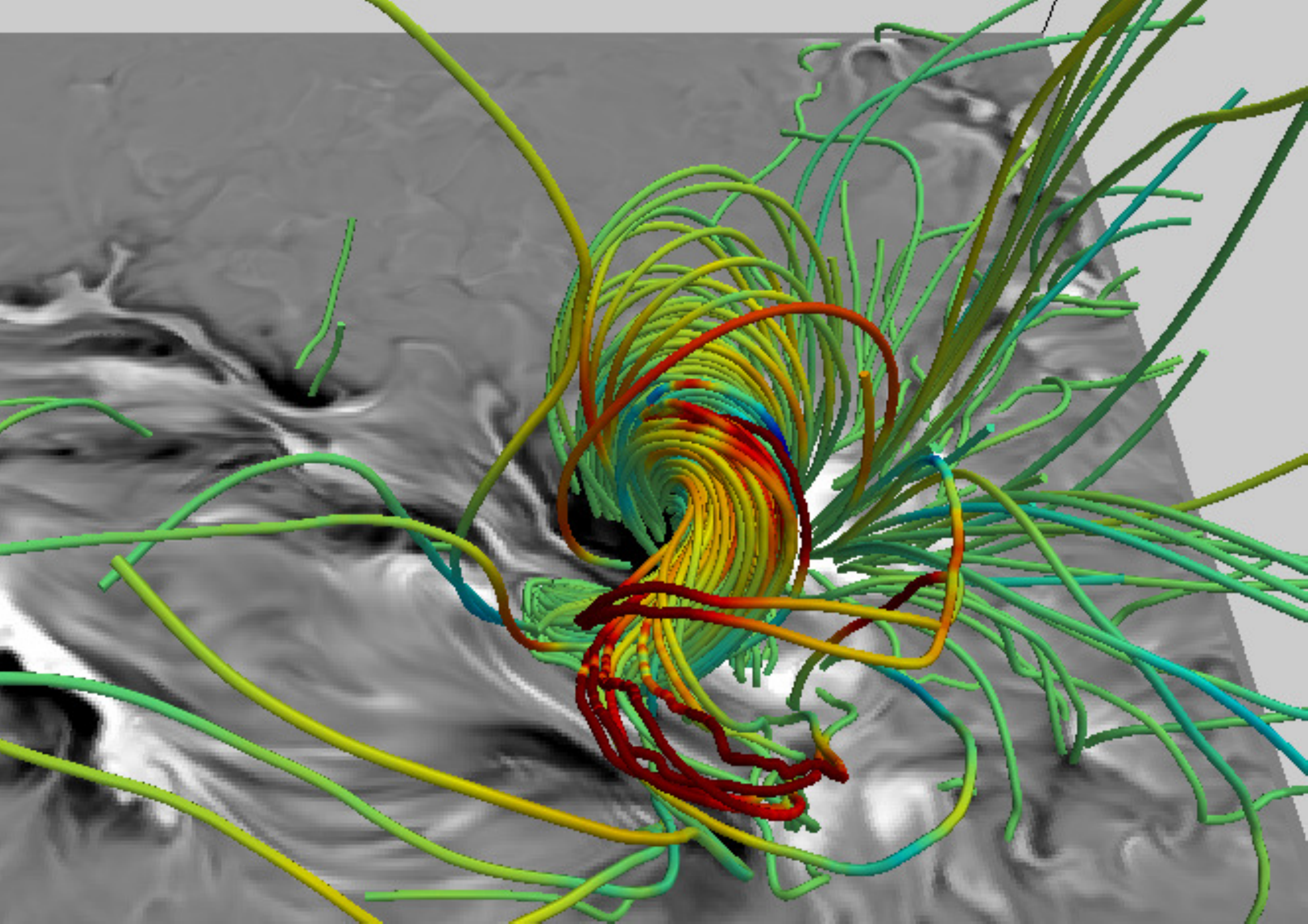}
\put(5,55) {\color{white}{(d) $t=240.2$ min}}
\put(40,30){\small\color{white}{FR-2}}
\put(100,-10){\includegraphics[width=0.24\textwidth]{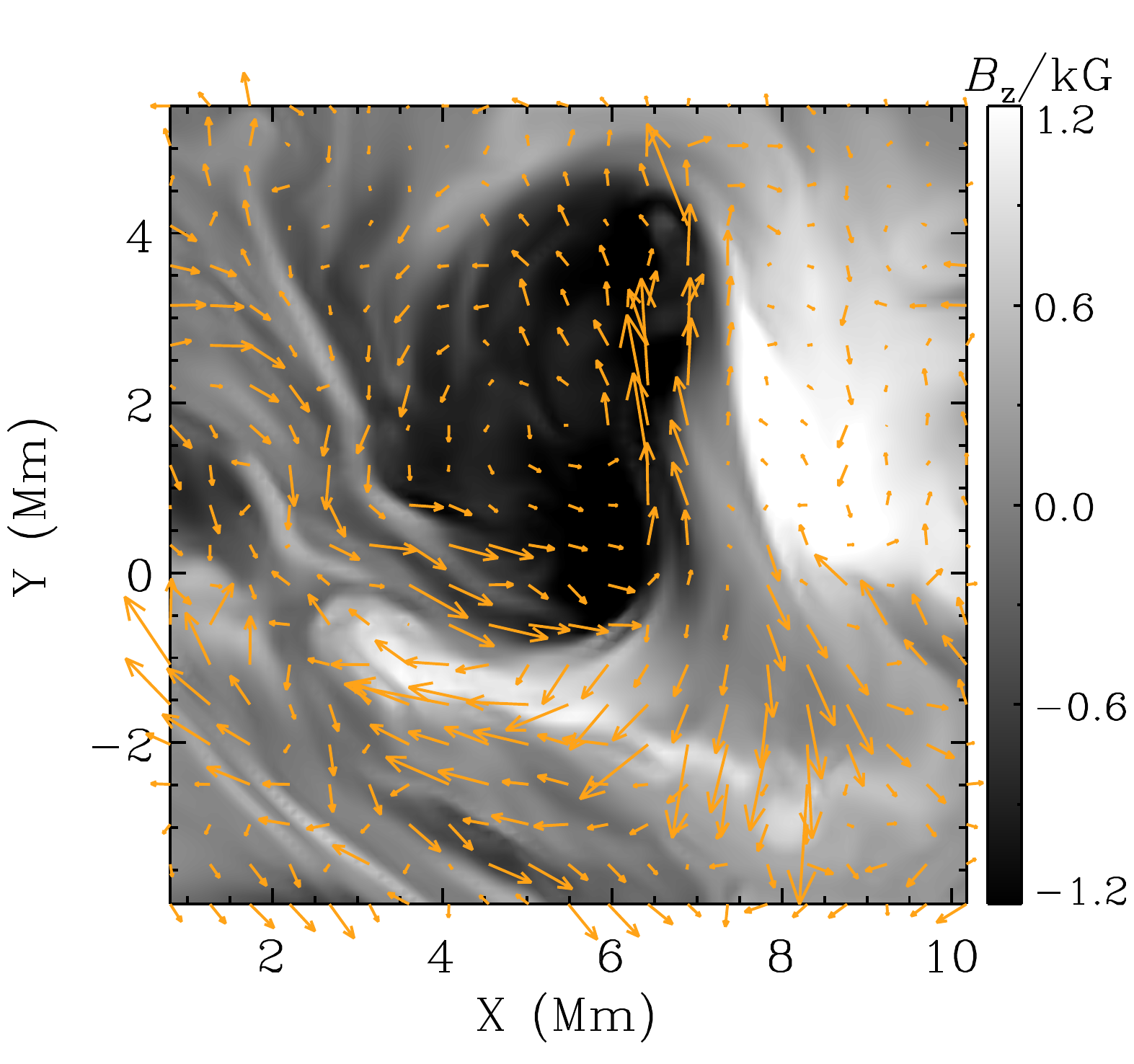}}
\put(120,80){\color{white} (e) $t=240.2$ min}
\end{overpic}}
\caption{\label{fig4} (a)--(d) Magnetic field lines coloured by the vertical velocity at times indicated to 
illustrate the formation and eruption of two flux ropes (FR). 
The field lines traced in (c) pass through the current-layer iso-surface (pink) 
with a value $J/B = 1/4\delta z$, where $cJ/4\pi$ is the current density, $c$ being the speed of light. (e) Horizontal velocity vectors 
in the $\delta$-sunspot region. The grey shaded surface represents the vertical 
magnetic field at $\textrm{z}=0$.  A fly-by animation of panel (c) also including 
the sub-surface structure of the $\delta$-sunspot is available at
\url{http://www.mn.uio.no/astro/english/people/aca/piyali/fluxemerge/fig5c.mp4}.
Also available is an animation of the time evolution of the magnetic field in this region at
\url{http://www.mn.uio.no/astro/english/people/aca/piyali/fluxemerge/fig5prl.mp4}.}
\end{figure}
The magnetic field lines at the key moments before and during the onset of 
flaring are visualised in Figure~\ref{fig4}. 
Panel (a) illustrates the formation of a flux rope (FR-1) due to several 
reconnections above the large opposite polarities during 
the first of these flares at $t=168.9$ min. Twenty-five minutes later
an inverse-$\mathcal{S}$-shaped flux-rope forms corresponding to a 
left-handed magnetic twist. The flux-rope erupts later, denoted as EFR-1 in panel (c) possibly because
it has been de-stabilised due to reconnections with the almost vertical magnetic field in the 
approaching positive polarity spot. The field lines, with red (blue) corresponding to upward (downward) velocity, 
clearly indicate the presence of a bipolar reconnection jet. These field lines pass 
through a current sheet, shown by a pink surface, 
with a thickness of $4\delta\textrm{z}$. 
The cusped reconnected field loops formed at this stage
develop into a new sigmoidal flux rope (FR-2)
at $t=240.2$ min (panel(d)). In panels (b)-(d), we note a {\it sigmoid-cusp-sigmoid} morphology transition 
over the region, also modelled by \cite{Chatterjee_Fan2013}, 
and often observed in the coronal soft X-rays above the source regions of 
homologous eruptions \citep{Gibson_etal2002}. 
Here, the inverse-$\mathcal{S}$-shaped flux rope is formed due to reconnections 
inside the current sheet followed by the 
shearing photospheric foot point motions, flux convergence and cancellation; supporting earlier 
observations by \cite{Martin_1990, Gaizauskas_etal1997} and simulations of 
\cite{vanBallegooijen_Martens1989, Martens_Zwaan2001}, rather than sub-surface flux tubes of Figure~\ref{fig1} (b) bodily
emerging into the photosphere.      
Figure~\ref{fig4} (e) shows strong shear at 
the polarity-reversal line between 
the spots which continuously pump twist and 
magnetic energy into the atmosphere in the form of a sustained Poynting flux 
$\sim 6 \times 10^{8}$ ergs s$^{-1}$ cm$^{-2}$ or $2\times10^{27}$ ergs s$^{-1}$ 
over the area of the box shown in Figure~\ref{fig2}. 
This value compares well with Figure 2 of \citep{Vemareddy2015} 
where they calculate the Poynting flux for C-class flares producing NOAA AR 11560 within an area $\sim 145\times105$ Mm$^2$. 
After this time, the vertical-component 
of the Poynting flux at $\textrm{z}=0$ is clearly 
dominated by the shearing foot-point motions rather than sub-surface flux emergence.
We also note the formation and eruption of a dense and cool filament-like structure 
above the $\delta$-sunspot region between $\textrm{z}=2.3 -  4.5$ Mm as shown in Figure~\ref{fig5} (a). 
At the beginning of the filament formation, 
the dense region grows, supported by the tilted field loops underneath. At a later time during the 
filament evolution, the tilted field loops develop dips into which the plasma flows.
This scenario is similar to $2.5$-dimensional simulation of funnel prominence formation from an 
arcade like geometry \citep{Xia_etal2012}
but is distinct from the cavity prominences where dipped or 
concave field lines supporting the denser plasma pre exist as part of an emerged 
flux rope in the corona \citep{Archontis_Hood2012}. The temperature inside the filament at $\textrm{z}=2.75$ Mm 
is $1.1 \times 10^5$ K which is one-third of the mean temperature in that layer. Hot 
plasma patches at 
$6.3\times10^5$ K exist close to the cooler filament. 
This phase separation of plasma into neighbouring regions of hot and cold has long been attributed to the onset of
thermal instability due to the radiative loss function \citep{Parker1953, Field1965}. Further, the evolution of these plasma condensations in long low-lying flux tubes is governed by the presence of both steady as well as impulsive heating just above the chromosphere \citep{Karpen_Antiochos2008}. 
We refer the readers to \cite{Mackay_etal2010} for a detailed review on filament formation and structure.
We define a quantity, $DE$, to describe the fractional density enhancement in the filament as $\rho/\bar{\rho}$, where 
$\bar{\rho}=\exp(\langle\ln\rho\rangle)$ with angular brackets denoting horizontal averaging. 
The region of plasma condensation
has an inverse-$\mathcal{S}$ shape and the maximum density is 116 times the ambient or 
$DE_{\textrm{\tiny max}}=116$. The total mass, at this stage, inside a volume bounded 
by the $DE = 10$ surface is $1.2\times10^{13}$ kg.
Panel (b) of Figure~\ref{fig5} shows a
part of the filament erupting at a speed of 50 km s$^{-1}$ along with the twisted flux rope depicted in 
panel (d) of Figure~\ref{fig4}. {The eruption speed is lower than observed, likely
because of the large viscosity and thermal diffusion used in this numerical simulation 
makes the conversion of the magnetic energy released 
to kinetic energy inefficient. The other possibility is the presence of a very weak pre-existing magnetic field incapable of 
confining the flux rope \citep{Fan2010, Aulanier_etal2010} long enough to build sufficient  
non-potential magnetic free energy in the system before its eruption. The magnetic free energy is a measure of the maximum energy available to drive eruptions. The larger the free energy, the faster may be the ejecta speeds. This may explain the absence of more energetic flares of class M and X in the simulated $\delta$-spot.}  
\begin{figure}
\label{fig5}
\begin{overpic}[width=0.43\textwidth]{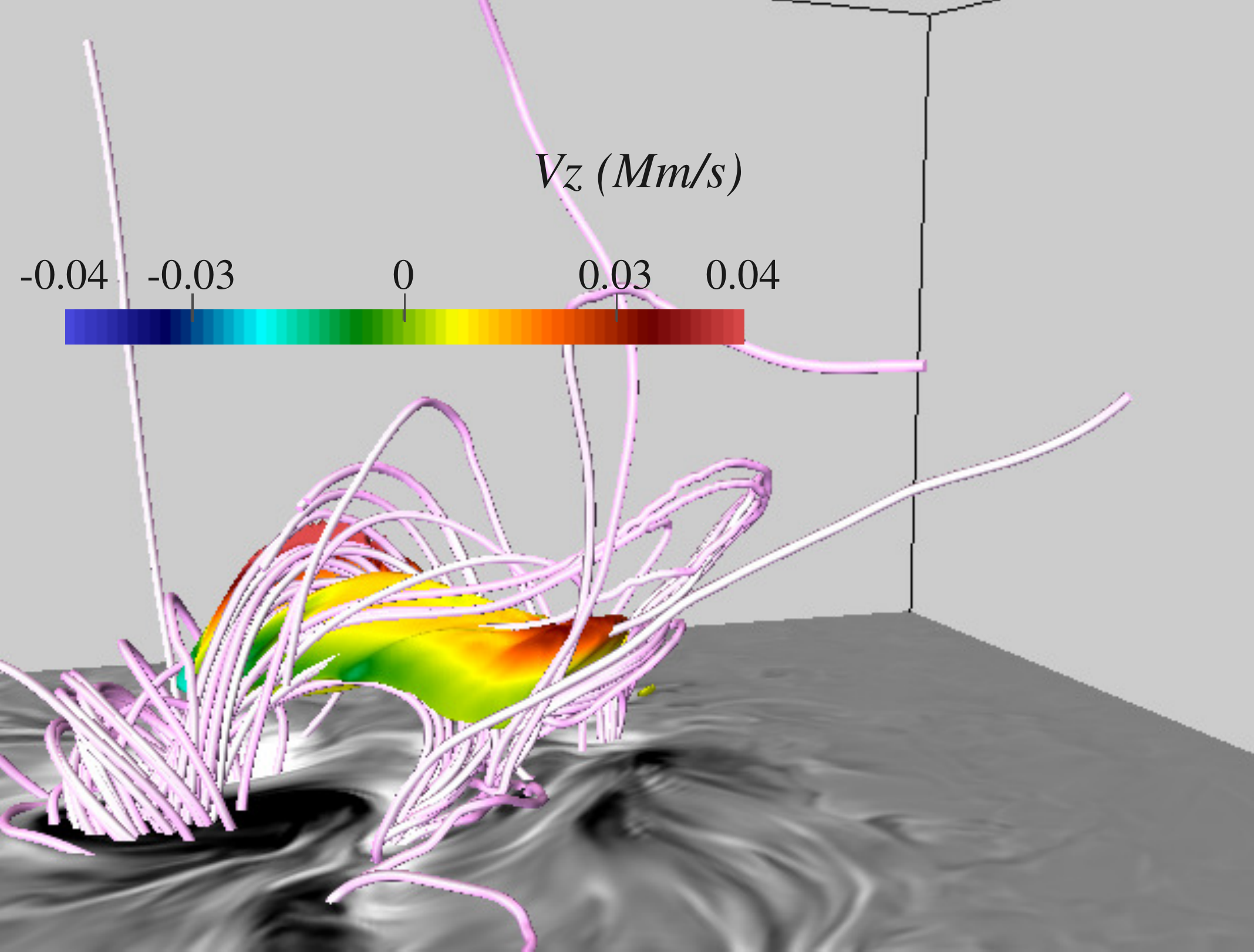}
\put(5,65) {(a) $t=240.2$ min}
\put(65,20){\includegraphics[width=0.16\textwidth]{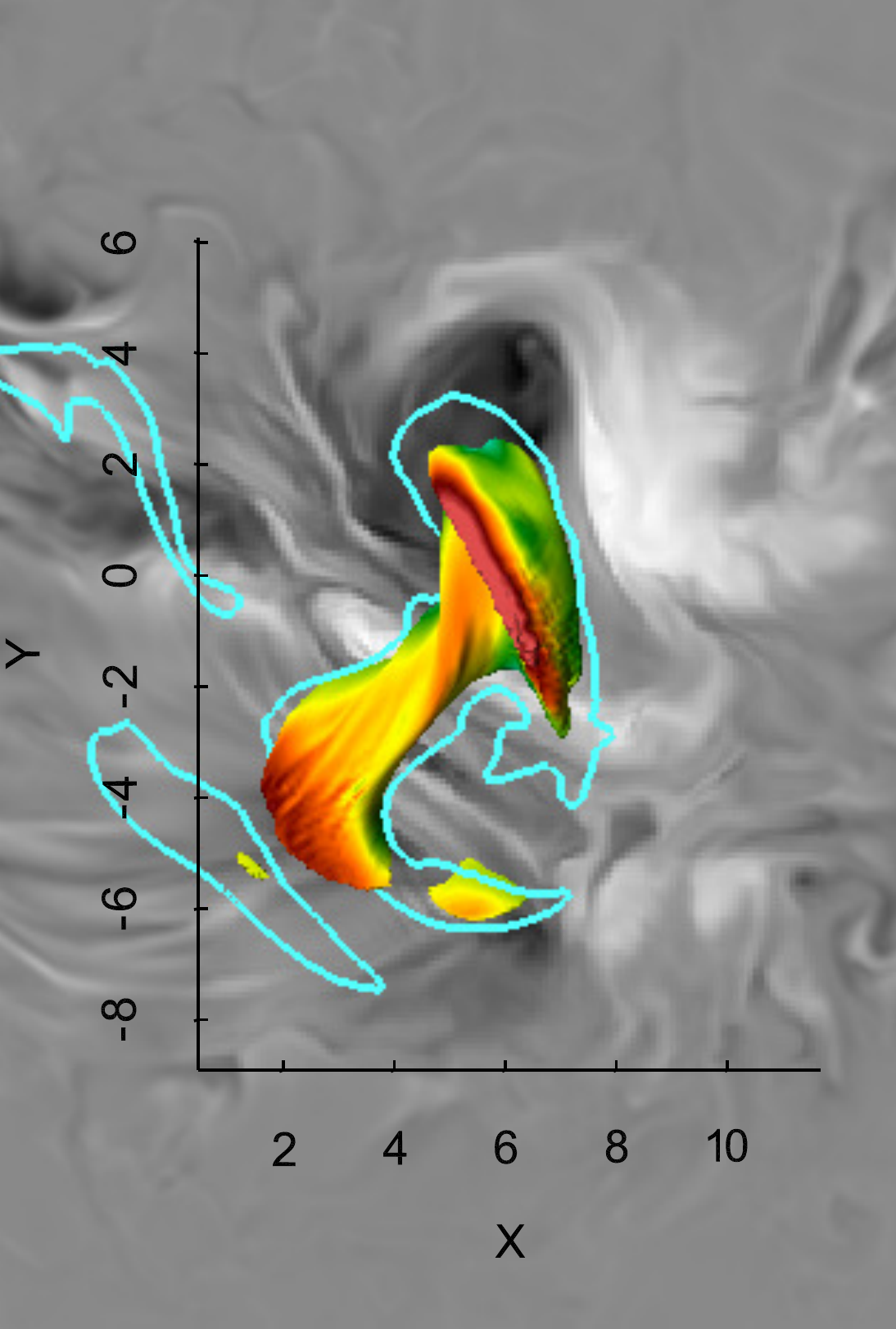}}
\end{overpic}
\begin{overpic}[width=0.43\textwidth,clip=true]{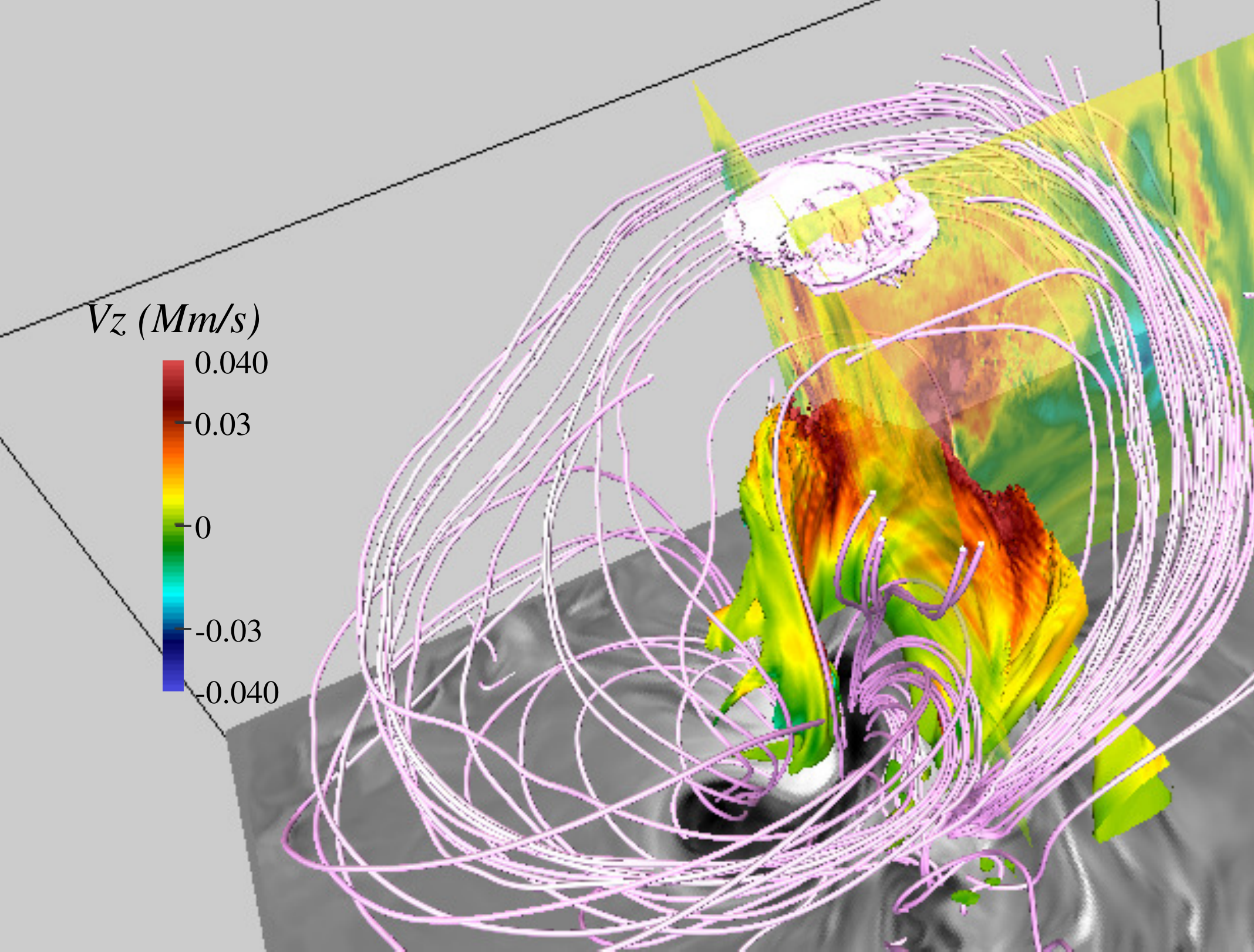}
\put(40,70) {(b) $t=263.6$ min}
\put(0,20){\includegraphics[width=0.16\textwidth,clip=true]{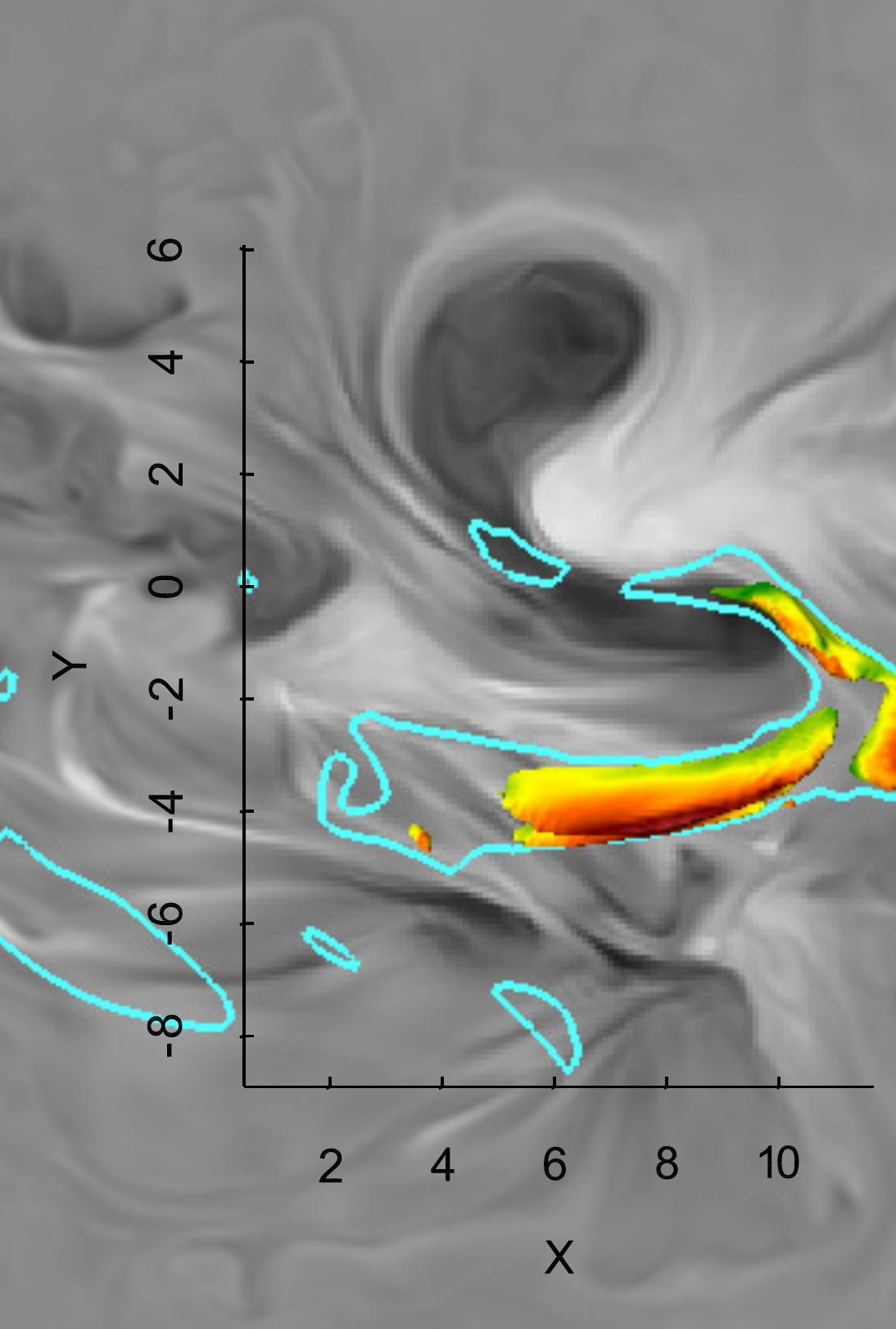}}
\end{overpic}
\caption{\label{fig5} Evolution of a dense filament over the $\delta$-sunspot. (a) $DE=21$ surface shaded by vertical velocity at 
the time indicated. (b) The $DE=6$ surface but 20 mins later when both the filament and the flux rope are erupting. 
The vertical planes are 
shaded by the vertical velocity to highlight the path of the erupting filament. The cloud in pink  
at the top has $DE=3.2$. The inset in both panels show the surface, $DE=21$, 
and line (cyan) contours with $DE=6$ at $\textrm{z}=2.3$ Mm superimposed on the top view of the photospheric magnetogram.}
\end{figure}

This numerical simulation started from a very primitive configuration, making no assumptions about the 
properties of sub-surface flux tubes,
demonstrates the formation
of a $\delta$-sunspot from
the collision of two or more young flux emerging regions developing in close vicinity. 
It is very similar to what is often seen in the 
solar photospheric magnetograms e.g., the widely studied active region with NOAA number 11158. 
However the two neighbouring regions in the 
vicinity occur not by mere chance, but emerge almost simultaneously as they are part of the same
initial subsurface 
structure.  
The collision leads to repeated flaring
which according to us causes
the pair to lock together throughout the evolution even though a major part of the 
component $\delta$-spots 
originate from topologically different flux tubes in the subsurface. This result validates the observational 
finding of \cite{Zirin_Liggett1987} and \cite{Zhongxian_Wang94} for $\delta$-sunspots from solar cycle~22. 
Another striking common 
feature of several observed $\delta$-sunspots  
e.g. NOAA AR 11158, 10488 \citep{Liu_Zhang2006} and 10808 \citep{Li_etal2007} and our simulation 
is the Yin-Yang structure
of the interpenetrating positive and negative $B_{\textrm{z}}$ in the late evolutionary phase 
(inset of Figure~\ref{fig5} (b)).  
Even though the treatment of the solar atmosphere is very simplified here,
we believe it captures the essential physics of magnetic flux emergence and evolution into a flaring $\delta$-sunspot. 
There is scope for improvement, for instance by including self consistent Ohmic heating of the corona, 
ionization, and detailed radiative transfer; this will be our future work.

\begin{acknowledgments}
 We thank two anonymous referees as well as O. V. S. N. Murthy for their comments which immensely helped improving the clarity of this paper. 
The research leading to these results has received funding from the European Research Council under the
European Union's Seventh Framework Programmes (FP7/2007-2013) / ERC grant agreement no 291058 and 
(FP7/2007-2013) /  F-CHROMA grant agreement no. 606862.
This research was also supported by the Research Council of Norway through the grant "Solar Atmospheric 
Modelling". The simulations were carried out on the NASA's Pleiades supercomputer 
under GID s1061. We have used the visualisation software Paraview for volume rendering and field line plotting.
\end{acknowledgments}
%
\end{document}